# Evidence for Anisotropic Vortex Dynamics and Pauli Limitation in the Upper Critical Field of FeSe$_{1-x}$Te$_x$

Daniel Braithwaite, Gérard Lapertot, William Knafo[1], and Ilya Sheikin[2]

SPSMS, UMR-E 9001, CEA-INAC/ UJF-Grenoble 1, 17 rue des martyrs, 38054 Grenoble, France

[1]LNCMI, UPR 3228, CNRS-UJF-UPS-INSA, 38042 Grenoble, France

[2] LNCMI, UPR 3228, CNRS-UJF-UPS-INSA, 31400 Toulouse, France

We have determined $H_{C2}(T)$ for FeSe$_{1-x}$Te$_x$ (x=0.52) single crystals using resistivity measurements at high static and pulsed magnetic field, as well as specific heat measurements up to 9T. We find that the significant anisotropy of the initial slope of $H_{C2}(T)$ determined from resistivity measurements, is not present when $H_{C2}$ is determined from the specific heat results. This suggests that the thermodynamic upper critical field is almost isotropic, and that anisotropic vortex dynamics play a role. Further evidence of anisotropic vortex dynamics is found in the behaviour in pulsed field. We also find that Pauli limiting must be included in order to fit the temperature dependence of $H_{C2,}$ indicating probably higher effective mass in FeSe$_{1-x}$Te$_x$ than in other Fe superconductors.

In the flurry of activity on the iron-arsenide superconductors after the discovery of superconductivity in the 1111[1] and 122[2] families, the discovery of superconductivity in tetragonal FeSe[3] is possibly an important piece in the puzzle of the mechanism of superconductivity in these systems. This system has the advantage of being a binary compound with a relatively simple structure, hopefully making theoretical approaches easier and more reliable than for more complicated systems. However to date it has proved difficult to grow single crystals showing evidence of homogeneous bulk superconductivity. Attention has therefore turned to some extent to the Te doped related system FeSe$_{1-x}$Te$_x$, which at close to 50% doping shows clear bulk superconducting properties[4]. The upper critical field is sensitive to the microscopic superconducting parameters, and its measurement is an important element in the research on the mechanism of high-$T_C$ superconductivity in these recently discovered iron superconductors. The critical fields of these systems are high, typically 40 – 60 T, meaning that to get the complete $H_{C2}$ curve pulsed magnetic fields must be used. A number of studies have now been published on the different families[5-12]. The general shapes of the obtained $H_{C2}$ curves are quite similar for the different families. Although globally $H_{C2}$ is weakly anisotropic, the slope at $T_C$ is found to display some anisotropy, being about a factor 2 larger when the field is applied perpendicular to the c-axis than when it is parallel. The anisotropy decreases at lower temperature, and the temperature dependences for the 2 orientations are therefore quite different. For H//c, in all studies $H_{C2}(T)$ shows a much less pronounced negative curvature, and even sometimes positive curvature. This behaviour is quite well described by a 2 band superconductivity model[5, 8], although the compatibility of the parameters with band structure calculations is not established[5]. Another important point is to establish if the paramagnetic Pauli limitation mechanism plays a role as this can have strong implications on the superconducting order parameter. The previous studies on the 1111 and 122 families showed that orbital limiting is sufficient to model the $H_{C2}(T)$ curves[5, 7, 12]. More recently studies on the 11 $FeSe_{1-x}Te_x$ system show that Pauli limiting seems to be present[9-11]. From the high-$T_C$ cuprates it is known that vortex dynamics can induce a difference between the critical field measured by resistivity, and the thermodynamic critical field, measured for example by specific heat[13]. In the $Ba_{0.6}K_{0.4}Fe_2As_2$ and $NdFeAsO_{1-x}F_x$ systems, specific heat measurements under field seem to confirm the anisotropy, although in the latter case the anomaly is quickly suppressed with field[14]. So far there has been no comparison between the critical fields obtained using the different techniques in $FeSe_{1-x}Te_x$. In this study we present resistivity measurements at high field to establish the full $H_{C2}(T)$ curves, combined with specific heat measurements up to 9T, in order to test the intrinsic nature of transition seen in resistivity.

Single crystals of $FeSe_{1-x}Te_x$ were grown with the Bridgman technique using a double wall quartz ampoule. The inside tube had a tipped bottom with a 30° angle and an open top. The inside wall of the outer ampoule was carbon coated to achieve the lowest possible oxygen partial pressure during the growth. Synthesis of the polycrystalline material was made prior to the growth in a similar quartz assembly at 800°C for 10 days. Material purities were Fe 4N6 (i.e. 99.996%) Se 5N and Te 5N. The Bridgman ampoule was inserted in a three zone gradient furnace (1000°/840°/700°) and lowered at a speed of 3 mm/h. At the end of the growth, temperature was lowered to room temperature at 50°C/h.

Superconductivity of the obtained crystals was characterized by resistivity, magnetization and specific heat measurements. Similarly to previous reports, resistive superconducting transitions were found for all concentrations, but evidence of bulk superconductivity from the transition in specific heat was only seen for values of x close to 0.5[4]. For this study we selected a crystal with x=0.52 in an attempt to possibly avoid or reduce the reported phase separation[4], though this was not characterised. A large piece (mass about 7 mg) was cleaved from the growth for specific heat measurements. A thin slice (m=0.8 mg) was cleaved off this piece for resistivity measurements. For these, a standard low frequency lock-in technique was used in static fields up to 28 T. A low temperature rotation mechanism allowed in-situ sample orientation. Measurements in pulsed fields up to 55 T were made by applying a high frequency (typically 40-60 kHz) ac current of about 1 mA and recording the sample voltage for the whole of the field pulse before extracting the resistance using a digital lock-in technique. In all cases the current was applied in the a-b plane. The sample was aligned by eye with an accuracy better than 2°. In the pulsed field set-up, the temperature was determined accurately when the sample was immersed in liquid helium for T<4.2 K, but a temperature gradient leading to errors of up to 0.5 K was estimated to exist at higher temperatures, where the sample was in exchange gas. The field was applied both parallel (H//c) and perpendicular (H//ab) to the c-axis. Specific heat measurements on the whole crystal were performed in a Quantum Design physical property measurement system (PPMS) for field up to 9T also applied parallel and perpendicular to the c-axis. Curves of resistivity and specific heat are shown in Fig. 1. The phonon contribution of the specific heat was estimated from the curve of a sample showing no transition in the specific heat, with a small renormalisation (less than 5%) at 20 K. The resulting estimation of the electronic part of the specific heat (insert Fig. 1) shows a clear superconducting transition, and a residual value of C/T that suggests that at least 80% of the sample is superconducting.

In Figure 2 resistive transitions are shown for pulsed fields (left plot) and static fields (right plot). In all cases up to the highest fields, zero resistivity was found in the superconducting state, and the transition remained relatively sharp. The obtained results for $H_{C2}$, taking the 50% resistivity criterion, are shown in figure 3. These results are similar to previous reports[6, 9] and confirm a significant anisotropy of the initial slope, and then a decreasing of the anisotropy at lower temperatures. In our case we see a small but clear inversion of the anisotropy at the lowest temperatures, where $H_{C2}$ (H//c) becomes larger than $H_{C2}$ (H//ab). At the lowest temperature measured (1.5 K) we find $H_{C2}$ values of 44 and 46 T for H//ab and H//c, respectively. The superconducting transition was also measured using specific heat for both directions of field up to 9 T. The transition remained well defined and showed little broadening (Fig. 4). The critical temperature defined as 50% of the increase of the electronic part of C/T coincides well at zero field with the 50% resistivity criterion. In fig 4 we show the comparison of the low field part of the phase diagram obtained from resistivity and specific heat. It appears clearly that the results of the specific heat for both field orientations agree with the resistivity results for H//ab, but not for H//c : In the specific heat no significant anisotropy is found between the two orientations.

Most of the early studies of the critical field in the Fe-As superconductors found that the $H_{C2}$ curves could be fitted taking into account only orbital effects[5, 7, 12]. The strong decrease of the anisotropy on decreasing temperature, and the sometimes observed positive curvature of $H_{C2}(T)$ for H//c could be described using a two band model with different anisotropies of the Fermi velocities for the two bands[5, 8]. Very recently studies on the $FeSe_{1-x}Te_x$ have shown that it is necessary to include Pauli paramagnetic limiting effects to describe the curves at least in this system[9-11]. From our specific heat measurements the intrinsic nature of the resistive $H_{C2}$ in $FeSe_{1-x}Te_x$ may be questionable, however so far the specific heat data is limited to low field so is not sufficient to fit the full $H_{C2}$ curve. We will therefore first look at the main conclusions that are obtained from the resistive $H_{C2}$. In Fig. 3 we show the best fits that we obtain using a weak coupling BCS clean limit model including orbital and Pauli limitation[15]. The hatched area indicates the transition width for H//ab. From our data we expect that for H//ab the resistive $H_{C2}$ is close to the thermodynamic one. It is obvious that, at least for H//ab, even though $H_{C2}$ exceeds quite significantly the standard Pauli limit, it is necessary to include Pauli limitation. We emphasize that the value found for the gyromagnetic factor (g=1) governing the Pauli limit has limited quantitative physical meaning: For example including strong coupling effects allows a reasonable fit to be obtained with g=2. The fact that Pauli limitation plays a larger role in $FeSe_{1-x}Te_x$ than in the other Fe based superconductors shows that the orbital $H_{C2}$ is higher in this system, indicating smaller Fermi velocities and higher effective mass than the other systems. We have performed a similar fit for H//c, although due to the large discrepancy between $H_{C2}$ measured from resistivity and specific heat, the significance of this is less clear. Interestingly in this case it is necessary to include an anisotropy of a factor 4 that in the Pauli limiting mechanism, revealed by the value of g=0.25 for H//c. The almost isotropic $H_{C2}$ at low temperature does not signify an isotropic Pauli limiting mechanism. Indeed even when Pauli limiting governs $H_{C2}$, the obtained value still depends on both the Pauli and orbital mechanisms to a varying degree.

The fact that $H_{C2}$ determined close to $T_C$ from the specific heat anomaly shows no anisotropy suggests that the thermodynamic upper critical field might be even more isotropic than usually believed, and that the apparent anisotropy determined from resistivity measurements could be mainly due to dissipation from vortex dynamics. This effect is well known in the high- $T_C$ cuprates where a second phase transition due to the melting of the vortex lattice can be defined[13]. In $FeSe_{1-x}Te_x$ the effect is not so large, indeed as can be seen in Fig. 3, the whole anisotropy is contained within the resistive transition width. The stronger broadening of the transition for H//c is sufficient to explain part of the anisotropy, and although a pertinent definition of the onset temperature is not easy to establish, the phase diagram using any definition of the onset temperature shows considerably less anisotropy. Further specific heat measurements at higher fields are planned.

Further evidence for anisotropic vortex dynamics was seen in the pulsed field measurement. When the field was applied along the c-axis, and the temperature was below 2 K, very strong peaks were seen in the sample voltage circuit at low field. In Fig. 5 we show the sample voltage for a small pulse up to about 2.5 T. This behavior was reproducible in two samples, and only found in this configuration on both samples. We interpret these peaks as the signature of vortex avalanches when the field is applied parallel to the c-axis. The signal would not be directly due to changes of the sample resistance, but rather to the voltage circuit acting as a pick-up to very fast changes of the sample magnetization. Similar effects have been seen in the magnetization of superconductors in the mixed state, mainly in strongly anisotropic or thin film material[16]. Of course the temperature range where the anomalous signal in the pulsed field measurement is found is far below the temperature where the anisotropy of the resistive upper critical field is large. We do not know how these two phenomena might be related, but both indicate anisotropic flux dynamics, and could stem from the same cause. It remains to understand why the flux dynamics should be so anisotropic if the superconductivity is isotropic. A possible explanation is the layered structure of the crystals, with possibly very different surface conditions for the two orientations. These results should stimulate further studies on flux dynamics in this and other Fe-based superconductors.

Acknowledgements

We are grateful to J.-P. Brison for help in fitting the $H_{C2}$ temperature dependences and critically reading the manuscript, and to B. Vignolle and C. Proust for help with the pulsed field measurements. Part of this work was supported by the French ANR project DELICE, and by Euromagnet II via the EU contract RII3-CT-2004-506239. One of us (I.S.) acknowledges support from the ANR project TETRAFER.

Figure Captions

Figure 1 : Low temperature resistivity (top) and specific heat (bottom) of a single crystal of FeSe$_{0.48}$Te$_{0.52}$. Inset shows the electronic part of C/T after subtraction of the phonon contribution (see text).

Figure 2 (color online) : Resistive transitions of the same single crystal of FeSe$_{0.48}$Te$_{0.52}$ measured in pulsed magnetic field applied H//ab (left) and static field H//ab (right).

Figure 3 (color online): Upper critical field of a single crystal of FeSe$_{0.48}$Te$_{0.52}$ determined from resistive measurements with the 50% criterion for both orientations of field. The error bars indicate the uncertainty of temperature working in exchange gas in the pulsed field set-up. The shaded area shows approximately the transition width for H//ab. Solid lines are calculations of H$_{C2}$ (T) using a simple weak coupling, one band model. A factor of more than 2 is found for the anisotropy of the initial slope (H'), and a factor 4 anisotropy must be included in the Pauli paramagnetic limiting mechanism (see text).

Figure 4 (color online) Left : Comparison between the transitions obtained by specific heat and by resistivity for both directions of field up to 9 T. Right : comparison of the low field part of H$_{C2}$ determined from resistivity and specific heat measurements. For H//ab there is a good agreement between the two techniques, but the anisotropy of the initial slope found in resistive measurements is not found when H$_{C2}$ is determined from the specific heat data.

Figure 5 (color online): Sample voltage versus time for a small (2.5 T) field pulse. The strong peaks in the signal for H//c were found on two samples, but only in the configuration H//c, and for temperatures below 2K. Curves have been shifted vertically for clarity. We interpret this phenomenon as flux penetration in the form of vortex avalanches, and evidence of anisotropic flux pinning. Bottom curve (right scale) shows field profile.

Figure 1

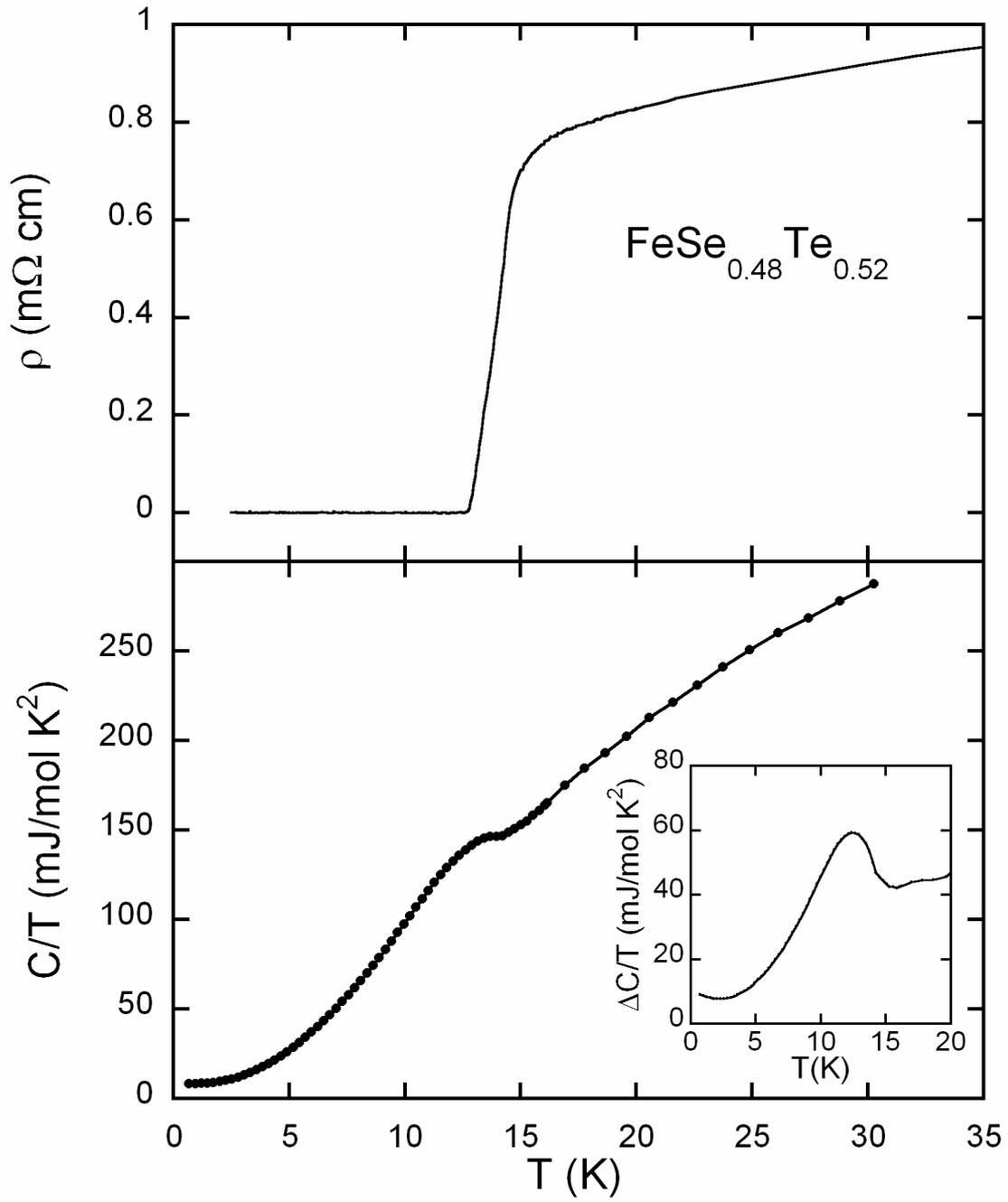

Figure 2

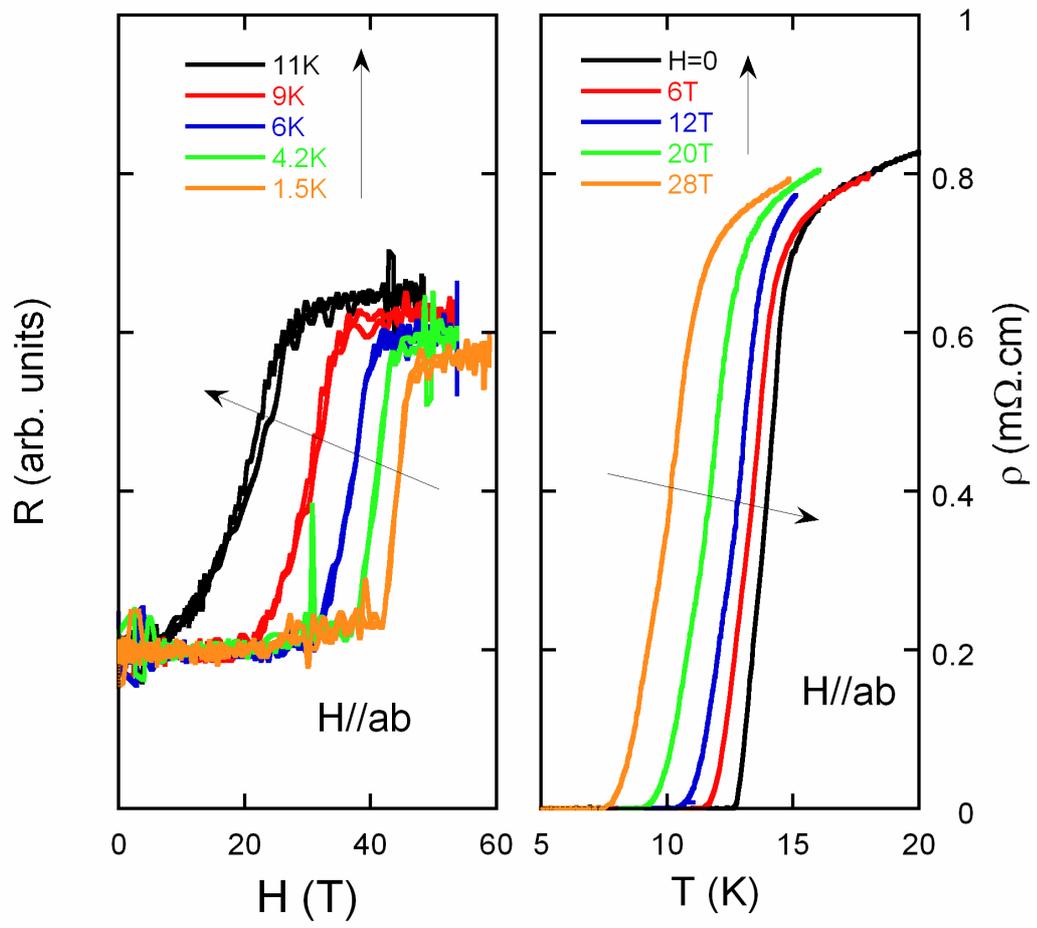

Figure 3

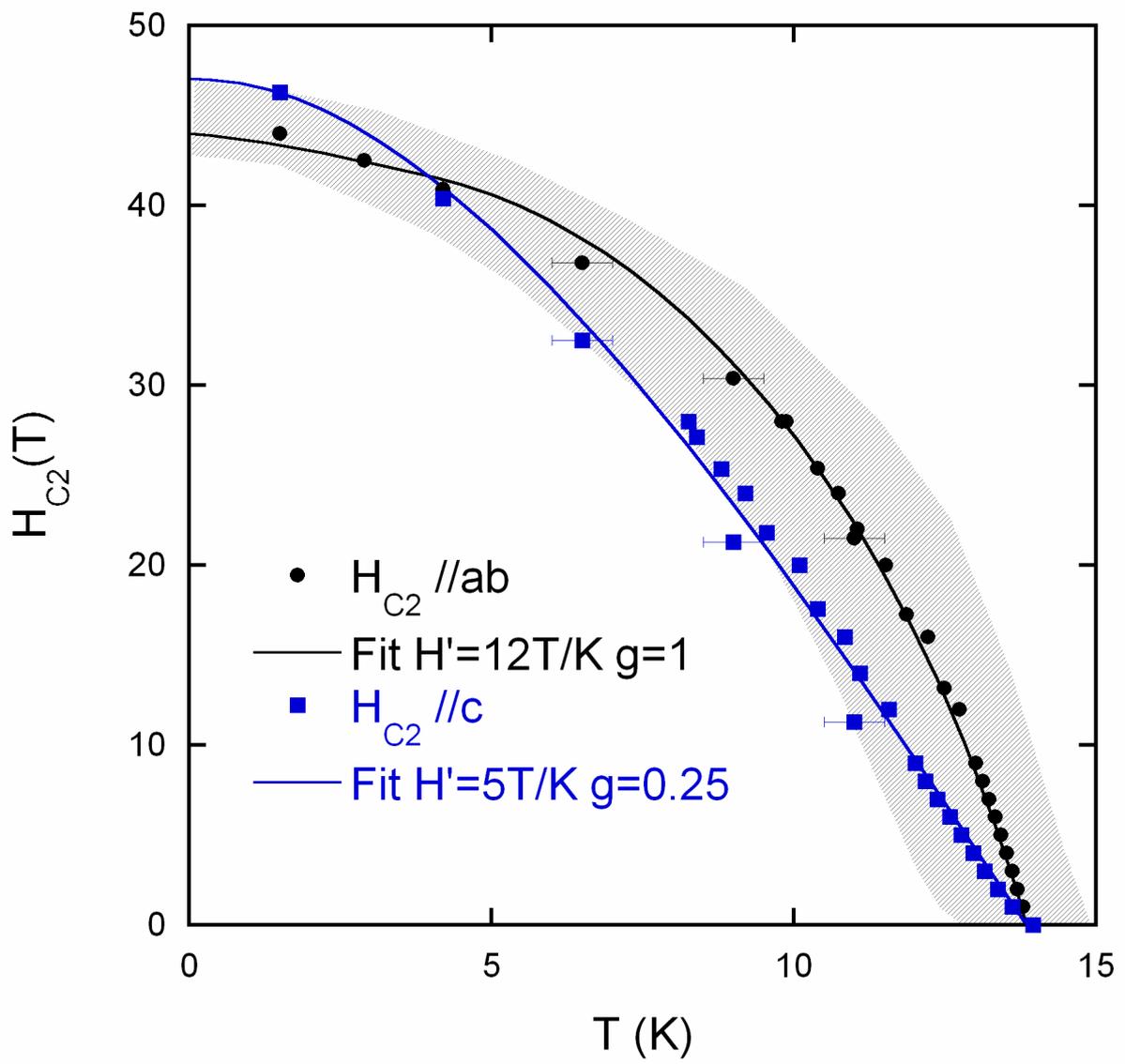

Figure 4

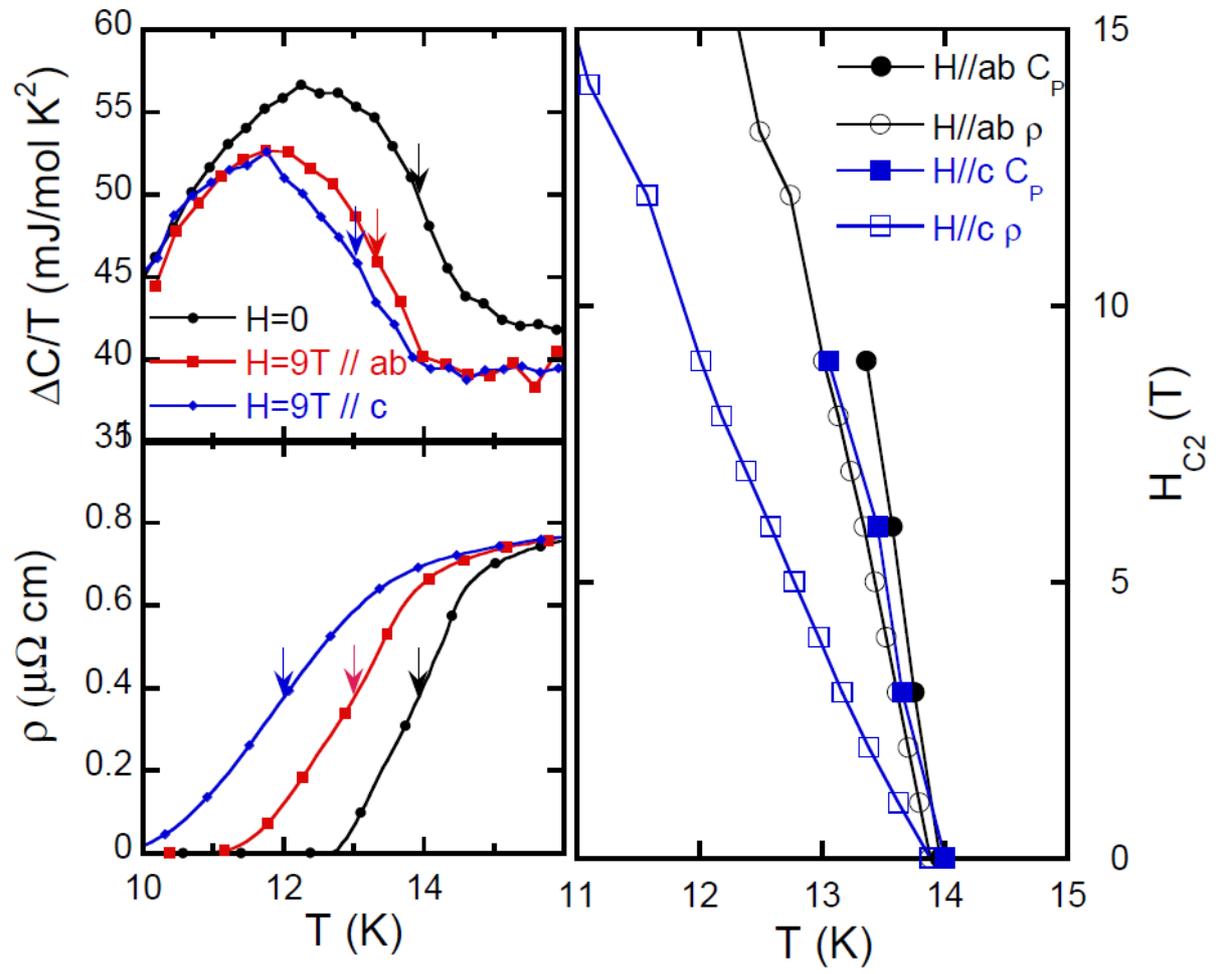

Figure 5

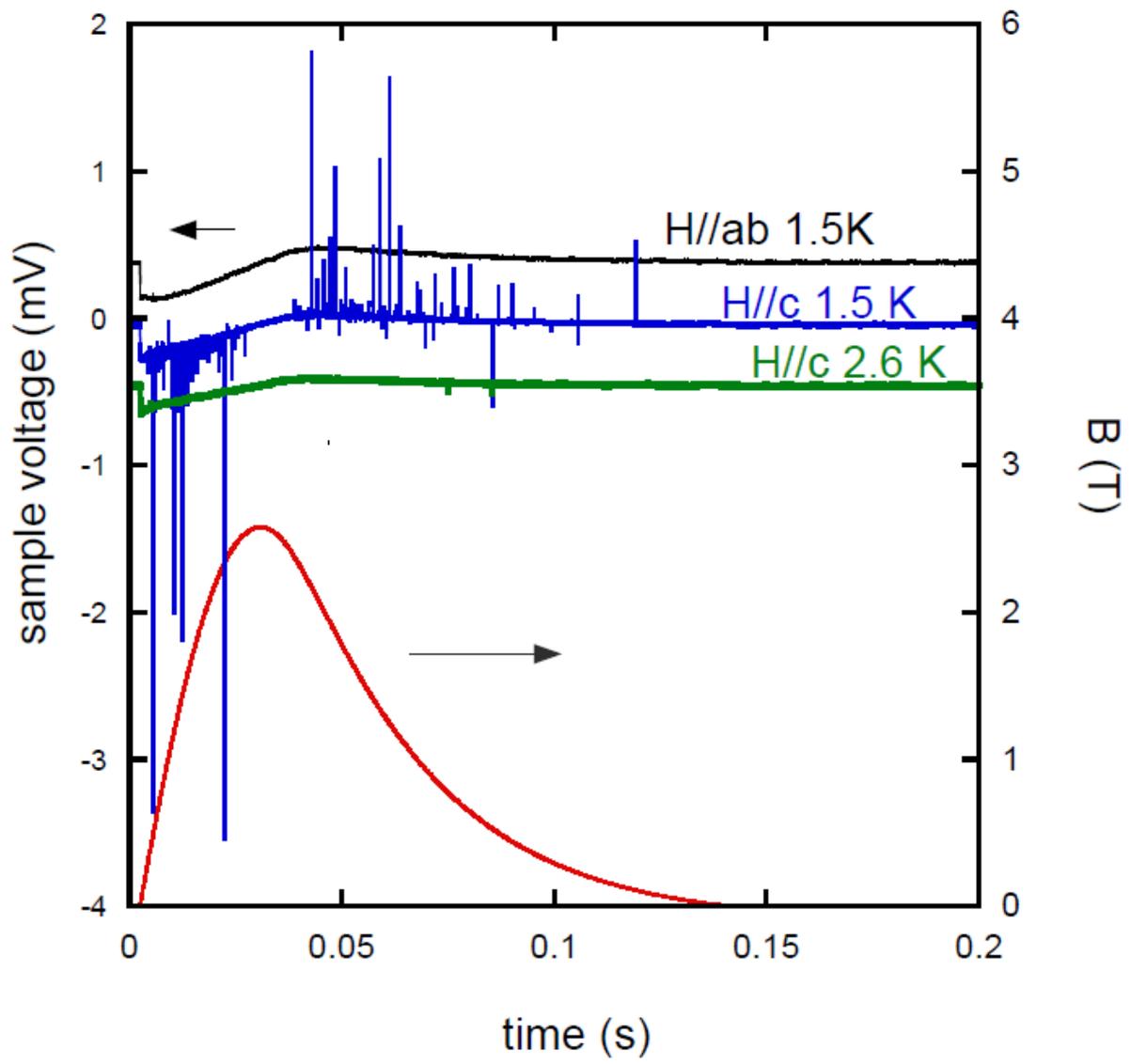